\documentclass[12pt]{article}
\usepackage{graphicx}
\usepackage{amsmath}
\usepackage{amssymb}
\usepackage{caption2}
\begin{document}
\bibliographystyle{prsty}
\begin{center}
{\large {\bf \sc{  Magnetic moment of the pentaquark
$\Theta^+(1540) $  with light-cone QCD sum rules }}} \\[2mm]
Zhi-Gang Wang$^{1}$ \footnote{Corresponding author;
E-mail,wangzgyiti@yahoo.com.cn.  }, Shao-Long Wan$^{2}$ and
Wei-Min Yang$^{2} $    \\
$^{1}$ Department of Physics, North China Electric Power University, Baoding 071003, P. R. China \\
$^{2}$ Department of Modern Physics, University of Science and Technology of China, Hefei 230026, P. R. China \\
\end{center}

\begin{abstract}
In this article, we study the magnetic moment of the pentaquark
state $ \Theta^+(1540)$ as diquark-diquark-antiquark
($[ud][ud]\bar{s}$) state in the framework of  the light-cone QCD
sum rules approach. The numerical results indicate the magnetic
moment of the pentaquark state $ \Theta^+(1540)$ is about
$\mu_{\Theta^+}=-(0.49\pm 0.06)\mu_N$.
\end{abstract}


\section{Introduction}

The observation of the new baryon $\Theta^+(1540)$ with positive
strangeness and minimal quark content $udud\bar{s}$ \cite{exp2003}
has motivated intense theoretical investigations
 to clarify its quantum numbers and to understand the
under-structures
\cite{jaffe03,karliner03,zhu03,matheus04,sugiyama04,eide04,Narison04,
Kanada04,takeuchi04,LatP,LatN,Shuryak04,Maltman04}. Through the
pentaquark state $\Theta^+(1540)$ can be signed to the top of the
antidecuplet $\overline{10}$ with isospin $I=0$, the spin and
parity have not been experimentally determined yet and no
consensus has ever been reached  on the theoretical side
\cite{Diakonov97, jaffe03,karliner03,
zhu03,matheus04,sugiyama04,LatP,LatN,Latt04,riska03,hosaka03,Carlson03p,
zhang03,carlson03n,wu03,Wang05}. The discovery has opened a new
field of strong interaction  and provides a new opportunity for a
deeper understanding of low energy QCD especially when multiquarks
are involved. The magnetic moments of the pentaquark states are
  fundamental parameters as their  masses, which have  copious  information  about the underlying quark
structures i.e. different substructures result in different
predictions, can be used to distinguish the preferred
configurations from various theoretical models and deepen our
understanding of the underlying dynamics.
 Furthermore, the magnetic moment of the pentaquark state $\Theta^+(1540)$ is an important ingredient
in studying the cross sections of the photo- or
electro-production, which can be used to determine the fundamental
quantum number of the pentaquark state $\Theta^+(1540)$, such
  as spin and parity \cite{HosakaPP,Nam04M}, and may be extracted from experiments eventually in the near
future.

 There have been several works on the magnetic moments of the pentaquark state
  $\Theta^+(1540)$ \cite{Huang04M,Zhao04M,Kim04M,Nam04M,Zhu04M,Inoue04M,
  Bijker04M,Goeke04M,Hong04M,Delgado04M,Wang05M,Wang05M2},
in this article, we take the point of view that the
$\Theta^{+}(1540)$ baryon is diquark-diquark-antiquark
($[ud][ud]\bar{s}$) state with the quantum numbers $J=\frac{1}{2}$
, $I=0$ , $S=+1$, and study its magnetic moment in the framework
of the light-cone QCD sum rules approach, which carries out the
 operator product expansion  near the light-cone $x^2\approx 0$
 instead of the short distance $x\approx 0$ while the nonperturbative
 matrix elements  are parameterized by the light-cone distribution amplitudes
 which classified according to their twists  instead of
 the vacuum condensates \cite{Balitsky86,Braun89}.

 The article is arranged as follows:   we derive the light-cone QCD sum rules for the magnetic moment of the
pentaquark state $\Theta^+(1540)$ in section II; in section III, numerical results; section IV is
reserved for conclusion.

\section{Light-cone QCD Sum Rules for the Magnetic Moment}
 We can study the magnetic moments of the baryons using
   two-point correlation functions  in an external
electromagnetic field, with
  vacuum susceptibilities introduced as parameters for
 nonperturbative propagation in the external field, i.e.
 the QCD  sum rules in the external field \cite{Ioffe84,Balitsky83}.
 As  nonperturbative vacuum properties, the susceptibilities can be
 introduced for both small and large momentum transfers  in the
 external fields . The alternative way
 is the light-cone QCD sum rules approach, which
  was firstly used to calculate the magnetic moments of
nucleons in Ref.\cite{Braun89}. For more discussions about the
magnetic moments of the baryons in the framework of the light-cone
QCD sum rules approach, one can consult Ref. \cite{Aliev}.

In the following, we write down  the two-point correlation
function $\Pi_{\eta}(p)$ in the framework of the light-cone QCD
sum rules approach \cite{matheus04,Zhu04M},
\begin{eqnarray}
 \Pi_{\eta} (p, q)&=&  i\int d^4 x \, e^{i p x}
\langle\gamma(q)|T\{\eta(x) \bar \eta(0) \}|0\rangle ,  \\
\eta(x)&=&\left\{t \eta_1(x) + \eta_2(x) \right\} , \\
\eta_1(x)&=&{1\over\sqrt{2}}\epsilon^{abc}\left\{\left[u_a^T(x) C
\gamma_5 d_b(x)\right] \left[u_c^T(x) C \gamma_5
d_e(x)\right]C\bar{s}^T_e(x) -
(u\leftrightarrow d)\right\} , \nonumber \\
\eta_2(x)&=&{1\over\sqrt{2}}\epsilon^{abc}\left\{\left[u_a^T(x) C
d_b(x)\right] \left[u_c^T(x) C  d_e(x) \right ]C\bar{s}^T_e(x) -
(u\leftrightarrow d)\right\}  . \nonumber
\end{eqnarray}
Here the $\gamma(q)$ represents the external electromagnetic field
with the vector potential $A_\mu (x)= \varepsilon_\mu e^{iq\cdot
x} $, the $\varepsilon_\mu$ is the photon polarization vector and
the field strength  $F_{\mu\nu}(x)=i( \varepsilon_\nu q_\mu
-\varepsilon_\mu q_\nu) e^{iq\cdot x}$. The $a$, $b$, $c$ and $e$
are color indexes,  the $C=-C^T$ is the charge conjugation
operator, and the $t$ is an arbitrary parameter. The constituents
$ \epsilon^{abc} u_b^T(x)C\gamma_5 d_c(x) $ and $ \epsilon^{abc}
u_b^T(x)Cd_c(x)  $ represent the scalar $J^P=0^+$ and the
pseudoscalar $J^P=0^-$ $ud $ diquarks respectively. They both
belong to the antitriplet $\bar{3}$ representation of the color
$SU(3)$ group and can cluster together with
diquark-diquark-antiquark type structure to give the correct spin
and parity for the pentaquark $\Theta^+(1540)$
$J^P={\frac{1}{2}}^+$ . The scalar diquarks correspond to the
$^1S_0$ states of $ud$ quark system. The one gluon exchange force
and the instanton induced force can lead to significant
attractions between the quarks in the $0^+$ channels
\cite{GluonInstanton}. The pseudoscalar diquarks do not have
nonrelativistic limit,  can be taken as  the $^3P_0$ states.

According to the basic assumption of current-hadron duality in the
QCD sum rules approach \cite{Shifman79}, we insert  a complete
series of intermediate states satisfying the unitarity   principle
with the same quantum numbers as the current operator $\eta(x)$
 into the correlation function in
Eq.(1)  to obtain the hadronic representation. After isolating the
pole terms of the lowest pentaquark  states , we get the following
result,
\begin{eqnarray}
\Pi_{\eta} (p, q) &=& -f_0^2\varepsilon^\mu\frac{{\not
p}+m_{\Theta^+}}{p^2-m_{\Theta^+}^2} [F_1(q^2)\gamma_\mu
+\frac{i\sigma_{\mu\nu}q^\nu }{2m_{\Theta^+}}F_2(q^2)]\frac{{\not
 p}+{\not q}+m_{\Theta^+}}{(p+q)^2-m_{\Theta^+}^2} +\cdots  \nonumber \\
 &=&-
\frac{f_0^2\left[F_1(q^2)+
F_2(q^2)\right]}{(p^2-m_{\Theta^+}^2)((p+q)^2-m_{\Theta^+}^2)}
\not\!p
\not\!\varepsilon (\not\!p+\not\!q) +\cdots \nonumber\\
&=&\Pi (p,
q)i\varepsilon_{\mu\nu\alpha\beta}\gamma_5\gamma^\mu\varepsilon^\nu
q^\alpha p^\beta+\cdots,
\end{eqnarray}
where
\begin{equation}
\langle 0| \eta(0) |\Theta^+ (p)\rangle =f_0 u(p) \, ,
\end{equation}
and the $F_1(q^2)$ and $F_2(q^2) $ are electromagnetic form
factors.  Here we write down only the double-pole term which
corresponds to the magnetic moment of the pentaquark state
$\Theta^+(1540)$ explicitly, and choose  the tensor structure
$\varepsilon_{\mu\nu\alpha\beta}\gamma_5\gamma^\mu\varepsilon^\nu
q^\alpha p^\beta$. The contributions concerning the excited  and
continuum states are suppressed after double Borel transformation,
and not shown explicitly for simplicity. From the electromagnetic
form factors $F_1(q^2)$ and $ F_2(q^2)$ , we can obtain the
 magnetic moment of the pentaquark state $\Theta^+(1540)$,
\begin{equation}
\mu_{\Theta^+}=\left\{ F_1(0)+F_2(0)\right\}
\frac{e_{\Theta^+}}{2m_{\Theta^+}}.
\end{equation}

The  calculation of the operator product expansion near the
light-cone $x^2\approx 0$ at the level of quark and gluon degrees
of freedom is
  straightforward and tedious, here technical details are neglected for
  simplicity. The photon can couple to the quark lines perturbatively and nonperturbatively,
  which results in  two classes of diagrams. In the first class of diagrams,
  the photon couples to
the quark lines perturbatively through the standard QED,
\begin{equation}
\langle 0|T\{q^a(x){\bar q}^b(0)\}|0\rangle_{F_{\mu\nu}}
=\frac{\delta^{ab}e_q}{16\pi^2x^2}\int_0^1
du\{2(1-2u)x_\mu\gamma_\nu
+i\epsilon_{\mu\nu\rho\sigma}\gamma_5\gamma^\rho x^\sigma
\}F^{\mu\nu}(ux).
\end{equation}
The second class of diagrams involve the nonperturbative
interactions of photons with the quark lines, which are
parameterized by  the photon light-cone distribution amplitudes in
stead of the vacuum susceptibilities.  In this article, the
following two-particle photon light-cone distribution amplitudes
are useful  \cite{Balitsky89,photon},
\begin{eqnarray}
\langle \gamma (q)|\bar q(x) \sigma_{\alpha\beta}q(0)|0\rangle
&=&i e_q \langle\bar qq\rangle \int_0^1 du \, e^{iu q x}
\left\{(\varepsilon_\alpha q_\beta - \varepsilon_\beta q_\alpha)
\{ \chi \varphi(u) + x^2 [(g_1(u) - g_2(u)]\} \right.
\nonumber\\
&&+\left. \{qx (\varepsilon_\alpha x_\beta - \varepsilon_\beta
x_\alpha) +
\varepsilon x (x_\alpha q_\beta - x_\beta q_\alpha)\} g_2(u)\right\} \, , \\
\langle \gamma (q)|\bar q(x) \gamma_\mu\gamma_5q(0)|0\rangle& =&
\frac{f}{4} e_q \epsilon_{\mu\nu\rho\sigma} \varepsilon^\nu q^\rho
x^\sigma \int_0^1 due^{iuqx} \psi(u) \, , \\
\langle \gamma (q)|\bar q(x) \gamma_\mu q(0)|0\rangle &=& f^{(V)}
e_q \varepsilon_\mu \int_0^1 due^{i uqx} \psi^{(V)}(u) \, ,
\end{eqnarray}
where the $\chi$ is the magnetic susceptibility of the quark
condensate, its values with  different theoretical approaches are
different from each other, for a short review,  one can see
Ref.\cite{Wang02}. The $e_q$ is the quark charge, $\varphi (u)$
and $\psi(u)$ are the  twist--2 photon light-cone distribution
amplitudes, while $g_1(u)$ and $g_2(u)$ are the twist--4
light-cone distribution amplitudes. The twist--3 photon light-cone
distribution amplitudes are neglected due to their small
contributions.

  Once  the analytical  results are obtained,
  then we can express the correlation functions at the level of quark-gluon
degrees of freedom into the following form through dispersion
relation,
  \begin{eqnarray}
  \Pi(p,q)= e_s\int_0^1 du \left\{\frac{1}{\pi}\int_{0}^{s_0}ds
  \frac{{\rm Im}[A(-s)]}{s-up^2-(1-u)(p+q)^2}+
  B(p,q) \right\}+\cdots\, ,
  \end{eqnarray}
where
\begin{eqnarray}
\frac{{\rm Im}[A(-s)]}{\pi}&=&
\frac{(5t^2+2t+5)s^4}{2^{12}5!4!\pi^8}-\frac{(5t^2+2t+5)s^3f
\psi(u)}{2^{13}5!\pi^6}+\frac{(7t^2-2t-5)s\langle
\bar{q}q\rangle^2}{2^{7}3^2\pi^4}\nonumber \\
&&-\frac{(7t^2-2t-5)\langle \bar{q}q\rangle^2f
\psi(u)}{2^{6}3\pi^2}+\frac{(5t^2+2t+5)s^2}{2^{15}4!\pi^6}\langle
\frac{\alpha_s GG}{\pi} \rangle  \nonumber \\
&&-\frac{(5t^2+2t+5)sf \psi(u)}{2^{14}3\pi^4}\langle
\frac{\alpha_s GG}{\pi} \rangle ,\nonumber \\
B(p,q)&=&-\frac{(5t^2+2t+5)\langle \bar{q}q\rangle^4}{2^{3}3^3
\left(-up^2-(1-u)(p+q)^2\right)^2}. \nonumber
\end{eqnarray}
From Eq.(10), we can see that due to the special interpolating
current (Eq.(2)), only the  $s$ quark has contributions to the
magnetic moment of the pentaquark $\Theta^+(1540)$, which is
significantly different from the results obtained in
Refs.\cite{Zhu04M,Wang05M}, where
 all the $u$, $d$ and $s$ quarks have contributions. In Refs.\cite{Zhu04M,Wang05M},
the  interpolating current $J(x)$ is used,
\begin{eqnarray}
J(x)={1\over \sqrt{2}} \epsilon^{abc} \left\{u^T_a(x) C\gamma_5
d_b (x)\right\} \{ u_e (x) {\bar s}_e (x) i\gamma_5 d_c(x) - d_e
(x) {\bar s}_e (x) i\gamma_5 u_c(x)  \} \, .
\end{eqnarray}
 Then we  make double Borel transformation with respect to the variables $p^2$ and
$(p+q)^2$ in Eq.(10),
\begin{equation}
{{\cal  B}}^{M_1^2}_{(p+q)^2} {{\cal  B}}^{M_2^2}_{p^2} {\Gamma
(n)\over [ m^2 -(1-u)(p+q)^2-up^2]^n }= (M^2)^{2-n} e^{-{m^2\over
M^2}} \delta (u-u_0 ) , \nonumber
\end{equation}
here $M^2=\frac{M_1^2M_2^2}{M_1^2+M_2^2}$ is the Borel parameter
and $u_0\equiv\frac{M_1^2}{M_1^2+M_2^2},
1-u_0\equiv\frac{M_2^2}{M_1^2+M_2^2}$. In this way the single-pole
terms  can be eliminated.  For more discussions about double Borel
transformation, one can consult Ref.\cite{Belyaev94}. Finally we
obtain the sum rule,
\begin{eqnarray}
\{ F_1(0)+F_2(0)\}f_0^2 e^{-\frac{m_{\Theta^+}^2}{M^2}}&=&-e_s AA,
\end{eqnarray}
\begin{eqnarray}
AA&=&\frac{(5t^2+2t+5)M^{12}E_4}{2^{12}5!\pi^8}-\frac{(5t^2+2t+5)M^{10}E_3f
\psi(u_0)}{2^{15}5\pi^6}\nonumber\\
&&+\frac{(7t^2-2t-5)\langle
\bar{q}q\rangle^2M^6E_1}{2^{7}3^2\pi^4}-\frac{(7t^2-2t-5)\langle
\bar{q}q\rangle^2f
\psi(u_0)M^4E_0}{2^{6}3\pi^2}\nonumber \\
&&+\frac{(5t^2+2t+5)M^8E_2}{2^{14}4!\pi^6}\langle \frac{\alpha_s
GG}{\pi} \rangle  -\frac{(5t^2+2t+5)f
\psi(u_0)M^6E_1}{2^{14}3\pi^4}\langle \frac{\alpha_s GG}{\pi}
\rangle \nonumber \\
&&-\frac{(5t^2+2t+5)\langle \bar{q}q\rangle^4}{2^{3}3^3 },
\nonumber \\
E_n&=&1-Exp\left(-\frac{s_0}{M^2}\right)\sum_{k=0}^{n}\left(\frac{s_0}{M^2}\right)^k\frac{1}{k!}\,
, \nonumber
\end{eqnarray}
here we have used the functions $E_n$ to subtract contributions
come from the excited states and continuum states. As the initial
and final states are the same pentaquark state $\Theta^+(1540)$.
It is natural to take the values $M_1^2=M_2^2=2M^2$ and
$u_0=\frac{1}{2}$.

If we replace the final $\gamma(q)$ state with the vacuum state in
Eq.(1),  we can obtain the sum rules for the coupling constant
$f_0$ \cite{matheus04},
\begin{equation}
f^2_0e^{-{m_{\Theta^+}^2\over M^2}}=BB,
\end{equation}
\begin{eqnarray}
BB &=&\frac{3(5t^2+2t+5)M^{12}E_5}{ 2^{11}7! \pi^8}
+\frac{(5t^2+2t+5)m_s\langle\bar{s}s\rangle M^{8}E_3}{2^{10}5!
\pi^6}+ \frac{(1-t)^2
M^{8}E_3}{2^{13}5!\pi^6}\langle \frac{\alpha_s GG}{\pi} \rangle \nonumber\\
& &+  \frac{(7t^2-2t-5)\langle \bar{q}q\rangle^2
M^{6}E_2}{2^{9}3^2\pi^4}- \frac{(5t^2+2t+5) m_s \langle
\bar{s}g_s\sigma \cdot G s\rangle  M^{6}E_2}{2^{14}3^2\pi^6}
\nonumber\\
&  &+\frac{(7t^2-2t-5) m_s \langle\bar{s}s\rangle \langle \bar{q}
q\rangle^2M^{2}E_0}{2^{6}3^2\pi^2} +\frac{(5t^2+2t+5)\langle
\bar{q} q\rangle^4}{6^3}\, .\nonumber
\end{eqnarray}
From above equations, we can obtain
\begin{eqnarray}
 \left\{F_1(0)+F_2(0)\right\} =-e_s\frac{
 AA} {BB} \,  .
\end{eqnarray}

\section{Numerical Results}
 In this article, we take the values of the parameter $t$ for the
interpolating current in Eq.(2) to be $t=-1$, which can give
stable mass i.e. $m_{\Theta^+}\approx 1540~MeV$  with respect to
the variations of the Borel mass $M^2$ in the considered interval
$M^2=(2-3)GeV^2$ in Ref.\cite{matheus04}.   The other parameters
are taken as $\langle \bar{s}s \rangle=0.8\langle \bar{u}u
\rangle$, $\langle \bar{u}u \rangle=\langle \bar{d}d \rangle=(-219
MeV)^3$, $\langle \bar{s}g_s\sigma\cdot G s \rangle=0.8\langle
\bar{s}s \rangle$, $\langle \frac{\alpha_s GG}{\pi}
\rangle=(0.33GeV)^4$,
 $m_u=m_d=0$ , $m_s=150MeV$,  $ \psi(u)=1$ and $f=0.028 \mbox{GeV}^2$
  \cite{Balitsky89,photon}. Here we have
neglected the uncertainties about
 the vacuum condensates, small variations of those condensates will not
 lead to larger changes about   the numerical
 values.
 The threshold parameter $s_0$ is chosen to  vary between $(3.6-3.8) GeV^2$ to avoid possible contaminations from
  higher resonances and continuum states.
 For $s_0=(3.6-3.8) GeV^2$, we obtain the values
 \begin{eqnarray}
 F_1(0)+F_2(0)&=&-(0.80\pm 0.10) \, , \nonumber\\
 \mu_{\Theta^+}&=& -(0.80\pm 0.10) \frac{e_{\Theta^+}}{2m_{\Theta^+}}\, , \nonumber \\
  &=& -(0.49\pm 0.06)\mu_N,
  \end{eqnarray}
where the $\mu_N$ is the nucleon  magneton.
 From the Table 1,  we can see that although  the numerical values for
 the magnetic moment of the pentaquark  state $\Theta^+(1540)$  vary with theoretical
 approaches, they are small in general;  our numerical results
are consistent with most of the existing values of theoretical
estimations in magnitude, however, with negative sign. In
previously work \cite{Wang05M,Wang05M2}, we observe that the sum
rules in the external electromagnetic field with different
interpolating currents can lead to very different predictions for
the magnetic moment, although they can both give satisfactory
masses for the pentaquark state  $\Theta^+(1540)$. The present
results show that the same interpolating current with different
approaches i.e. the QCD sum rules in the external fields and the
light-cone QCD sum rules, can result in different predictions in
magnitude with the same sign \cite{Zhu04M,Wang05M,Wang05M2}.  When
the experimental measurement of the magnetic moment of the
pentaquark state $\Theta^+(1540)$
  is possible in the near future, we might  be able  to test the theoretical
   predictions of the magnetic moment and  select
the preferred quark configurations.
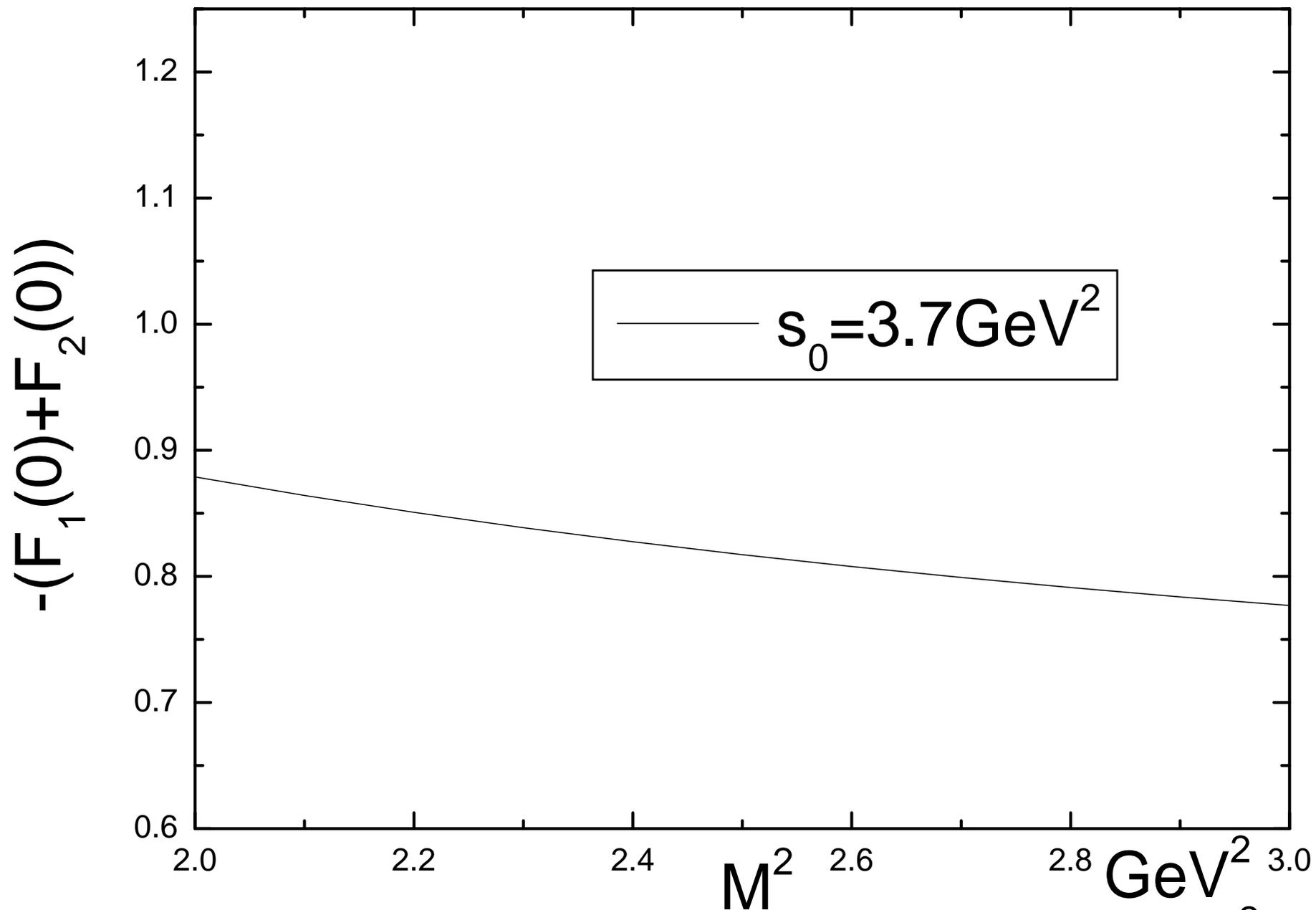
     In the region $M^2=(2-3)GeV^2$, the sum rules for $F_1(0)+F_2(0)$
     as functions of the Borel parameter $M^2$ is plotted in Fig.1 for $s_0=3.7 GeV^2$ as an example.

  \begin{table}[ht]
         \caption{\label{Tablechi}The values of $\mu_{\Theta^+}$ (in unit of $\mu_N$)}
         \begin{center}
         \begin{tabular}{c||c}
         \hline\hline
         Reference          & $\mu_{\Theta^+}$ \\
                            &  $(\mu_N)$  \\ \hline\hline
     \cite{Huang04M}         &   0.12 $\pm$ 0.06        \\ \hline
    \cite{Zhao04M}     &   0.08 $\sim$ 0.6        \\ \hline
     \cite{Kim04M} & 0.2$\sim$0.3     \\ \hline
     \cite{Nam04M}        & 0.2$\sim$0.5         \\ \hline
      \cite{Zhu04M}              & 0.08 or 0.23 or 0.19 or 0.37           \\ \hline
      \cite{Inoue04M}              & 0.4          \\ \hline
      \cite{Bijker04M}             & 0.38          \\ \hline
      \cite{Goeke04M}             &-1.19 or -0.33         \\ \hline
     \cite{Hong04M}             &0.71 or 0.56          \\ \hline
      \cite{Delgado04M}            &0.362          \\ \hline
\cite{Wang05M}            &0.24$\pm$0.02          \\ \hline
            \cite{Wang05M2}         &-(0.134$\pm$ 0.006)           \\ \hline
            This Work          &-(0.49$\pm$ 0.06)           \\ \hline\hline
         \end{tabular}
         \end{center}
         \end{table}
\section{Conclusion }

In summary, we have calculated  the magnetic moment of the
 pentaquark state  $\Theta^+(1540)$ as diquark-diquark-antiquark
($[ud][ud]\bar{s}$) state in the framework of  the light-cone  QCD
sum rules approach. The numerical results are consistent with most
of the existing values of theoretical estimations in magnitude,
however, with negative sign, $\mu_{\Theta^+}= -(0.80\pm 0.10)
\frac{e_{\Theta^+}}{2m_{\Theta^+}}
  = -(0.49\pm 0.06)\mu_N$ .
In  previously work \cite{Wang05M,Wang05M2}, we observe that the
sum rules in the external electromagnetic field with different
interpolating currents can lead to very different predictions for
the magnetic moment, although they can both give satisfactory
masses for the pentaquark state  $\Theta^+(1540)$. The present
results show that the same interpolating current with  different
approaches  i.e. the QCD sum rules in the external fields and the
light-cone QCD sum rules, can result in  different predictions in
magnitude with the same sign \cite{Zhu04M,Wang05M,Wang05M2}. The
magnetic moments of the baryons are fundamental parameters as
their masses, which have copious information  about the underlying
quark structures, different substructures can lead to
 very different results. The width of the pentaquark state $\Theta^+(1540)$ is so
narrow, the
 small magnetic moment may be extracted from photo-production
experiments eventually in the near future, which may be used to
distinguish the preferred configurations from  various theoretical models, obtain
more insight into the relevant degrees of freedom and deepen our understanding about  the underlying
dynamics that determines the properties of the exotic pentaquark  states.

\section*{Acknowledgment}
This  work is supported by National Natural Science Foundation,
Grant Number 10405009,  and Key Program Foundation of NCEPU. The
authors are indebted to Dr. J.He (IHEP) and Dr. X.B.Huang (PKU)
for numerous help, without them, the work would not be finished.

\end{document}